\documentclass[fleqn,usenatbib]{mnras}

\usepackage{newtxtext,newtxmath}

\usepackage[T1]{fontenc}

\DeclareRobustCommand{\VAN}[3]{#2}
\let\VANthebibliography\thebibliography
\def\thebibliography{\DeclareRobustCommand{\VAN}[3]{##3}\VANthebibliography}

\usepackage{graphicx}	
\usepackage{amsmath}	
\usepackage{amssymb}	
\usepackage{siunitx}
\usepackage{comment}
\usepackage{multirow}
\usepackage{physics}
\usepackage{threeparttable}

\title[The origin of long soft lags in the hard-intermediate state]
{The origin of long soft lags 
and the nature of the hard-intermediate state in black hole binaries
}

\author[T. Kawamura et al.]{
Tenyo Kawamura,$^{1,2}$\thanks{E-mail: tenyo.kawamura@gmail.com}
Chris Done,$^{3,2}$
and Tadayuki Takahashi$^{2,1}$
\\
$^{1}$Department of Physics, University of Tokyo, Bunkyo, Tokyo 113-0033, Japan\\
$^{2}$Kavli Institute for the Physics and Mathematics of the Universe (WPI), University of Tokyo, Kashiwa, Chiba 277-8583, Japan\\
$^{3}$Centre for Extragalactic Astronomy, Department of Physics, Durham University, South Road, Durham DH1 3LE, UK
}

\date{Accepted XXX. Received YYY; in original form ZZZ}

\pubyear{2023}

\begin{document}
\label{firstpage}
\pagerange{\pageref{firstpage}--\pageref{lastpage}}
\maketitle

\begin{abstract}
Fast variability of the X-ray corona in black hole binaries can
produce a soft lag by reverberation, 
where the reprocessed thermalized disc photons lag behind the illuminating hard X-rays. 
This lag is small, and systematically decreases with increasing mass accretion rate towards the hard-soft transition, consistent with a decreasing truncation radius between the thin disc and X-ray hot inner flow. 
However, the soft lag suddenly increases dramatically just before the spectrum becomes disc-dominated (hard-intermediate state). 
Interpreting this as reverberation requires that the X-ray source distance from the disc increases dramatically, potentially consistent with switching to X-rays produced in the radio jet. 
However, this change in lag behaviour occurs without any clear change in hard X-ray spectrum, and before the plasmoid ejection event which might produce such a source (soft-intermediate state). 
Instead, we show how the soft lag can be interpreted in the context of propagation lags from mass accretion rate fluctuations. These normally produce hard lags, as the model has radial stratification, with fluctuations from larger radii modulating the harder spectra produced at smaller radii. However, all that is required to switch the sign is that 
the hottest Comptonized emission has seed photons which allow it to extend down in energy below the softer emission from the slower variable turbulent region from the inner edge of the disc. 
Our model connects the timing change to the spectral change, and gives a smooth transition of the X-ray source properties from the bright hard state to the disc-dominated states. 
\end{abstract}




\section{Introduction}
\label{sec:intro}

Black hole binaries show distinct `states' in their accretion flow spectra, switching from 
a sum of blackbodies (soft spectrum) to a Compton-dominated (hard spectrum). 
The soft spectral state is well described by the Shakura-Sunyaev optically thick, geometrically thin disc (\citealt{Shakura_1973, Pringle_1981}), but the nature and geometry of the accretion flow in the hard state is much more controversial (\citealt{Yuan_2014}). 
Most of the spectral and variability properties can be well described by models where the inner thin disc is replaced by X-ray hot plasma (truncated disc, hot inner flow models), where the truncation radius of the thin disc decreases as the source luminosity increases,  until the disc reaches the last stable circular orbit (e.g. \citealt{Done_2007}). 

However, there is a persistent challenge to these models from the detection of the iron line and reflected continuum produced by the hard X-ray source illuminating the disc. 
The line is broadened by special and general relativistic effects, the extent of which depends on the radial extent of the disc (\citealt{Fabian_1989}). 
Many of the lines detected in bright hard states are so broad as to require that the disc extends down to the last stable circular orbit rather than being truncated (e.g. \citealt{Wang-ji_2018, Buisson_2019}).
This may be due to the assumption hardwired into current reflection models all that the flux comes from above, whereas there is clearly also intrinsic flux from the disc, which will change the 
baseline ionisation state. Whatever the cause, the discrepancy in disc inner radius has motivated studies of alternative geometries, where the X-ray plasma is a compact corona on the black hole spin axis, illuminating an untruncated disc (lamppost: e.g. 
\citealt{Miniutti_2004}), or is extended along the spin axis (jet corona: e.g. \citealt{Kara_2019}). 

One way to distinguish between these different model geometries is to use the additional information present in the fast variability.
The variable hard X-rays illuminate the disc, and the reflected emission should follow this variability but with a light travel time lag encoding the mean distance of the X-ray source from the disc (\citealt{Fabian_2009}, see also the review by \citealt{Uttley_2014}). This is difficult to see in data above 2~keV as the iron line/reflected continuum is not dominant in the spectrum at these energies. However, the thermalized, reprocessed emission can make a much larger contribution to the intrinsic disc emission below 1~keV. Identifying this reverberation signal requires an instrument with high effective area in this soft bandpass, so it was discovered with XMM-Newton data \citep{DeMarco_2015}, 
though now the \textit{NICER} instrument is giving better constraints (\citealt{DeMarco_2021}, hereafter dM21, \citealt{Wang_2021}, hereafter W21). There is a very clear pattern in the data that the reverberation lag decreases as the source softens (dM21, W21), consistent with the truncated disc model prediction, but also consistent with a source on the spin axis, where the mean height decreases as the source softens \citep{Kara_2019}. 

These reverberation (soft) lags do not exist in isolation. 
They are formed against a background hard lag, which is important to model simultaneously in order to isolate the reverberating component (\citealt{Mastroserio_2018, Mastroserio_2021}). 
These hard lags depend on the frequency of the fluctuation: slow fluctuations show a larger lag than faster fluctuations (\citealt{Miyamoto_1988, Nowak_1999}).
Hard lags are generally explained in the truncated disc framework as fluctuations generated in and propagating down through a radially extended hot accretion flow \citep{Ingram_2011, Ingram_2012}. 
Slow fluctuations are generated at large radii where the emitted spectrum is softer. These propagate down and modulate faster fluctuations generated at smaller radii where the spectrum is harder. 
Full physically based models of this can fit all the variability seen in the 2--40~keV data of MAXI~J1820+070 in a bright hard state (\citealt{Kawamura_2023}, hereafter K23). 
Nonetheless, it may be possible to modify the model to fit with the alternative untruncated disc/compact corona, though this geometry is now challenged by new X-ray polarisation data \citep{Krawczynski_2022}. 

Whatever the bright hard state geometry, there is a characteristic, quite sudden, increase in soft lag as the source softens at the end of the bright hard state into the hard-intermediate state (dM21, W21, \citealt{Wang_2022}). 
These authors interpreted this lag in the same framework of reverberation as used for the much smaller lags seen in the bright hard state. 
This requires an equally sudden jump in mean source distance, so they speculated that this was linked to the discrete radio ejection which happens at the end of the hard-intermediate state (\citealt{fender_2009, miller-jones_2012, Homan_2020}).
In this model, the increase in lag then marks the point at which the X-ray source transitions from being close to the black hole to being dominated by the base of the jet at a distance of hundreds of gravitational radii $R_{\mathrm{g}}=GM_{\mathrm{BH}}/c^2$ away from the black hole on its spin axis (dM21, W21). 
A similar jump is also seen in the lags at the QPO frequency seen in transitions of the luminous transient GRS1915+104, which is similarly interpreted as a large scale geometry change of the X-ray source \citep{Mendez_2022}.
However, this is not easily compatible with the observation that the radio jet ejections happen at the end (rather than at the start) of the hard-intermediate state (\citealt{Homan_2020, Bright_2020, Wood_2021}).
Nor is there any evidence for any correlated dramatic change in the coronal hard X-ray source, or the reflected spectrum from the disc, as shown in Section~2. 

Here, we revisit the \textit{NICER} and \textit{Swift} data of MAXI~J1820+070 in the hard-intermediate state, modelling both the spectrum and variability together to show that there is an alternative interpretation of the soft lag in a propagating fluctuation framework rather than reverberation.
The key difference which turns the hard lag from propagation into a soft lag is linked to the change in X-ray emission properties, where the disc truncates at much smaller radii so the turbulent inner edge emission peaks at higher energies. Fluctuations from this region propagate down to modulate the inner Comptonization component which extends the emission to higher energies, generating the classic hard lag. 
However, this can also generate a soft lag if the inner Comptonisation component has seed photons which extend its emission into the 0.5--1~keV band, below the peak emission from the turbulent inner disc edge. 
Our model connects the timing change to the spectral change, and gives a smooth transition of the X-ray source properties from the bright hard state to the hard-intermediate state. 


\section{Observation and data reduction}
\label{sec:data}

\begin{figure}
	\includegraphics[width=1.\columnwidth]{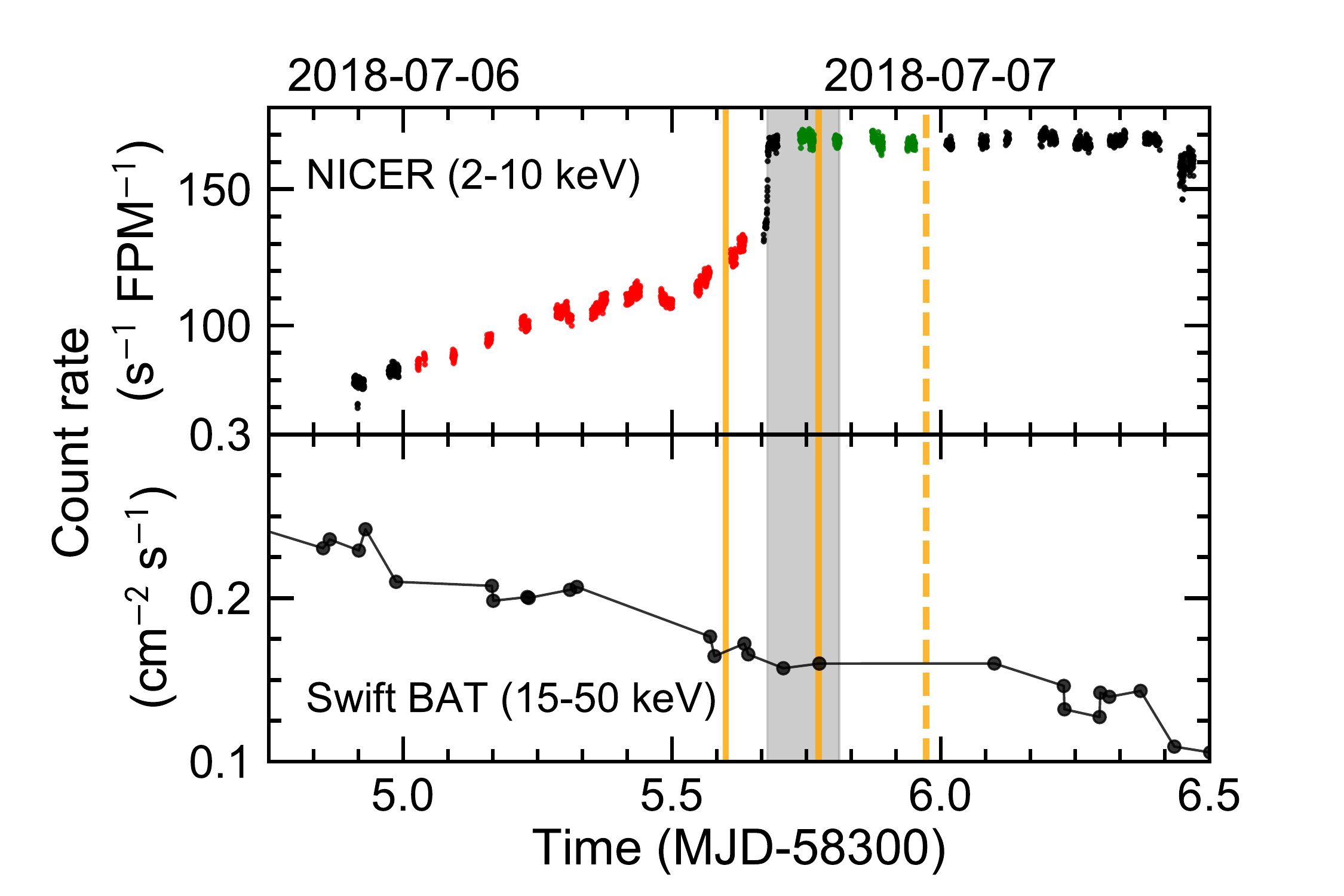}
	\caption{Light curves of MAXI~J1820+070. 
    The top panel shows the 2--10~keV \textit{NICER} data from Obs IDs 1200120196, 1200120197, and 1200120198 on 16~s time binning.
    The HIMS (red) is defined as the first 54~ks of the Obs ID 1200120197, while the SIMS (green) is as the rest of the data but the abrupt change of the count rate.
    The lower panel shows the 15--50~keV \textit{Swift} BAT data. 
    The vertical solid orange lines mark different estimates for the radio ejecta launch times, while the orange dashed line shows the time of the radio flare peak (\citealt{Homan_2020, Wood_2021}).
    The grey-shaded area marks the period when the Type-B QPO are observed (\citealt{Homan_2020}).
    }
    \label{fig:j1820_lightcurves}
\end{figure}

We use \textit{NICER} and \textit{Swift} BAT data of MAXI~J1820+070 around the hard-soft state transition.
We follow \cite{Kawamura_2022} to extract energy spectra and light curves from the \textit{NICER} data.
We split the light curves into segments of 256~s with 1/128~s time bins and calculate power spectra and cross spectra also as in \cite{Kawamura_2022}.
Orbital light curves from the \textit{Swift} BAT data are directly obtained \footnote{\url{https://swift.gsfc.nasa.gov/results/transients/weak/MAXIJ1820p070/}} (\citealt{Krimm_2013}), while energy spectra are extracted in the same way as in \cite{Shidatsu_2019}. 

Fig.~\ref{fig:j1820_lightcurves} (upper panel) shows the \textit{NICER} 2--10~keV count rate per focal plane module (FPM) around the hard-soft state transition spanning Obs IDs of 1200120196, 1200120197, and 1200120198. 
We follow dM21 and select the hard-intermediate state (HIMS) data from the first part of Obs ID 1200120197 (red markers) before the jump in count rate which marks the transition from HIMS to the soft intermediate state (SIMS), and the appearance of Type-B QPOs (grey shaded area, from \citealt{Homan_2020}).
The second part of Obs ID 1200120197 (green markers) is in the SIMS. 
The successive observation of Obs ID 1200120198 shows that the source is very stable in 2--10~keV flux in the SIMS.

There are contemporaneous data from \textit{Insight-HXMT} around this time
\citep{Peng_2023}, but the closest dataset starts just after the transition into the SIMS, so we cannot use these to extend the spectral/timing HIMS analysis to higher energies.
Instead, we use the \textit{Swift} BAT all-sky data. 
Fig.~\ref{fig:j1820_lightcurves} (lower panel) shows the 15-50~keV count rate from this instrument. 
This does not have the clear jump in flux seen at lower energies, there is only a gradual decline throughout this time period. 
We extract the \textit{Swift} BAT spectral data corresponding to the HIMS, and to the SIMS. 

We also mark on Fig.~\ref{fig:j1820_lightcurves} the times from \citet{Homan_2020} for the radio flare peak (orange dashed line) and their estimate for the start time of this radio event (orange solid line, within the grey shaded region: see \citealt{Bright_2020}). 
An alternative estimate for the time of the radio ejection event is that of \cite{Wood_2021} of MJD $55305.60\pm 0.04$ (leftmost orange solid line), i.e., 2 hours before the transition from Type-C QPOs to Type-B QPOs.
These two estimates for the launch time of the discrete ejecta span the time range of the end of the HIMS/beginning of the SIMS X-ray transition. Plainly they do not extend back as far as the start of the HIMS.  
This clearly challenges the model of dM21 and W21, where the corona changes from a compact region close to the black hole to being at hundreds of $R_{\mathrm{g}}$, associated with the jet ejection event. The jet ejection event happens 
around the HIMS/SIMS transition, so cannot be responsible for the change in lag behaviour which happens before this, at the transition from the bright hard state to the HIMS. 


\section{Evolution of spectral-timing properties}
\label{sec:spectral_timing_properties}

\begin{figure}
	\includegraphics[width=1.\columnwidth]{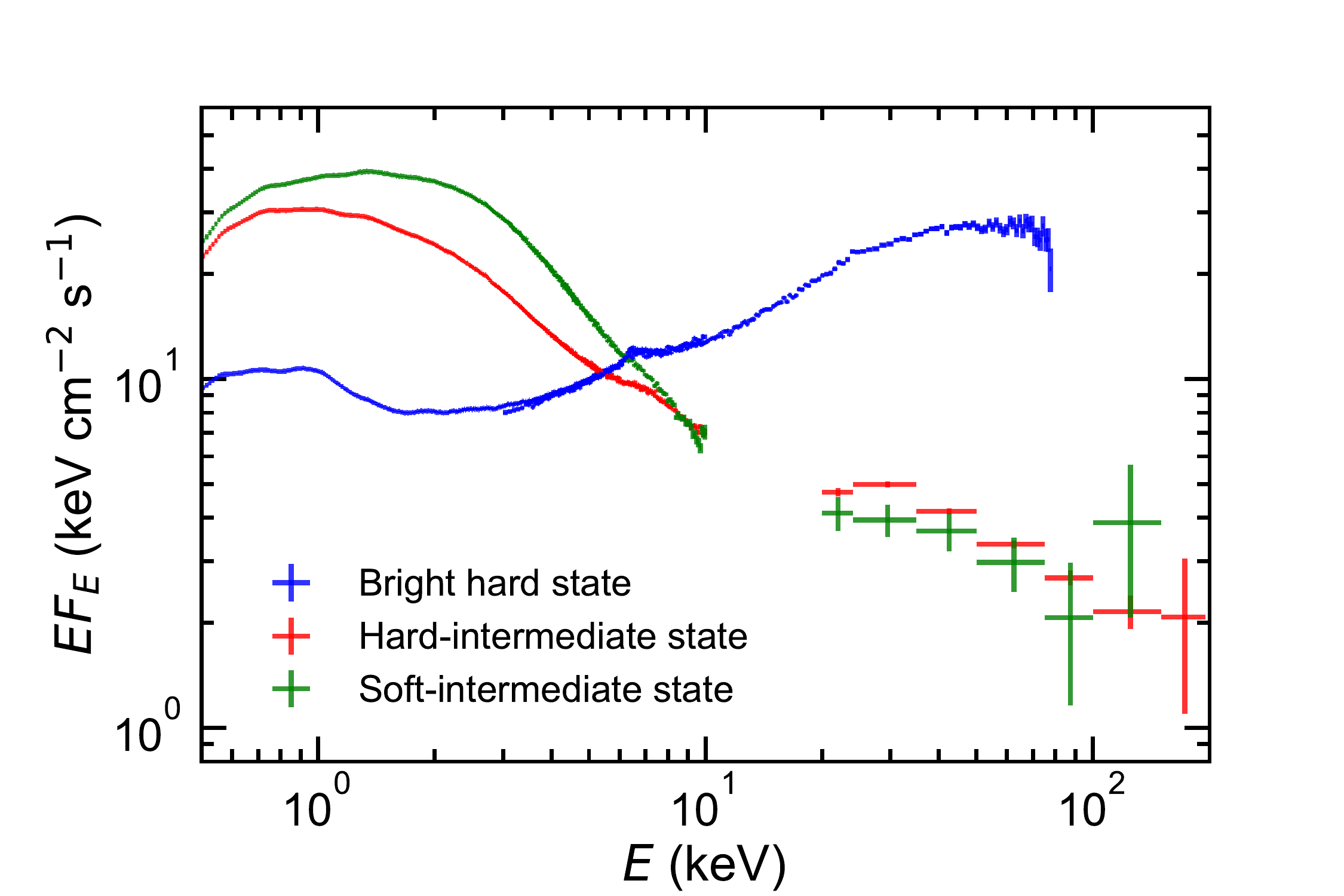}
		\caption{Comparison of absorption corrected energy spectra in the bright hard state (\textit{NICER} and \textit{NuSTAR}: blue; \citealt{Kawamura_2022}), hard-intermediate state (\textit{NICER} and \textit{Swift} BAT: red), and soft-intermediate state (\textit{NICER} and \textit{Swift} BAT: green). 
		The galactic absorption of $N_{\mathrm{H}}=1.4 \times 10^{21}\,\si{cm^{-2}}$ is assumed (\citealt{Zdziarski_2021}).
		It is clear that the main change from HIMS to SIMS is the shape of the disc component, not the high-energy tail.
    }
    \label{fig:nicer_j1820_hims_spectrum}
\end{figure}

\begin{figure*}
	\includegraphics[width=0.85\linewidth]{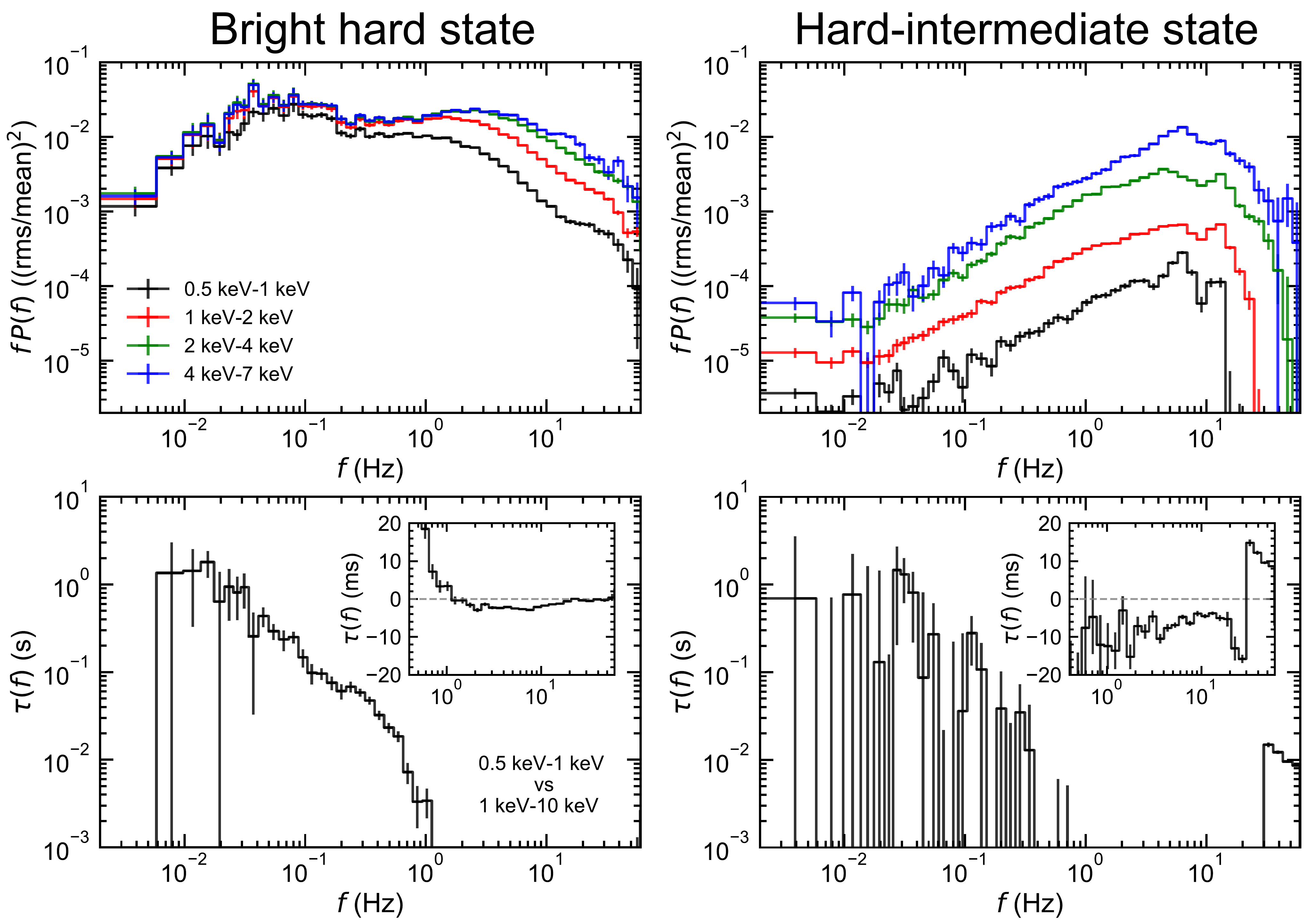}
    \caption{Comparison of fast variability properties of MAXI~J1820+070 in the bright hard state (Obs ID: 1200120106) and hard-intermediate state (Obs ID: 1200120197, first 54~ks).
    Top: Power spectra for 0.5--1~keV (black). 1--2~keV (red), 2--4~keV (green). and 4--7~keV (blue).
    Bottom: Time-lag spectrum between 0.5--1~keV and 1--10~keV.
    The inset shows the time lags on a linear scale.
    }
    \label{fig:ni106_psd_lag}
\end{figure*}

Fig.~\ref{fig:nicer_j1820_hims_spectrum} shows the absorption corrected spectra for the HIMS (red) and SIMS (green), together with a bright hard state (blue: \textit{NICER} and \textit{NuSTAR}: \citealt{Kawamura_2022}) for comparison. 
These are some of the best broadband spectra of the HIMS and SIMS obtained to date due to the
combination of low galactic column 
which allows the soft band to be studied in detail and source brightness which allows the high energy source spectrum to be extracted from the \textit{Swift} BAT all-sky monitor data.
The comparison clearly shows that the transition is associated with a change in the soft component shape, not the high-energy coronal tail. 
The soft component in the HIMS appears quite strongly Comptonized rather than having a classic disc shape with a Wien rollover at energies above the peak disc temperature. 
The SIMS is more disc-like (though we note it is still difficult to fit with classic disc models). 

This association of the HIMS/SIMS transition with a change in the disc rather than the high-energy Compton tail is again challenging for the model of dM21 and W21, where they associate the increasing lags with the corona changing from a compact region close to the black hole to being at hundreds of $R_{\mathrm{g}}$, associated with the jet ejection event. 
It is hard to imagine how such a dramatic change in coronal structure has such a little impact on the coronal emission spectrum or luminosity. 
There is also still a strong, broad iron line residual in the HIMS spectrum, as well as in the bright hard state. 
If the location of the corona had changed from close to the black hole to a much more distant source, then the ionisation parameter, $\xi=L/(nR^2)$, of the X-ray illuminated disc should drop dramatically. 
The line should also be much less affected by relativistic smearing as the illumination from a large-scale height source is more like $r^0$ than $r^{-3}$ (or even steeper with light bending). 
Yet the line residual appears similarly (or perhaps even more strongly) broadened in the HIMS as in the bright hard state. 
These disconnects motivate our approach to find another way to interpret the long soft lag in the context of the same model geometry used for the bright low/hard state. 

Since the long soft lag already emerged before the HIMS/SIMS transition associated with discrete jet ejection events and the appearance of Type-B QPO (Fig.~\ref{fig:j1820_lightcurves}), it is natural to try to smoothly connect the HIMS spectra and timing with the bright hard state spectra and timing.
Fig.~\ref{fig:ni106_psd_lag} shows the bright hard state (left) and HIMS (right) power spectra in different energy bands (upper), and the lags between the 0.5--1~keV and 1--10~keV bands (lower), corresponding to the spectral data (bright hard state: blue and HIMS: red) in Fig.~\ref{fig:nicer_j1820_hims_spectrum}.
Our previous work (\citealt{Kawamura_2022}, K23) was able to fit the bright hard state spectrum and variability (power spectra in different energy bands and lags) in a physically based model where the accretion disc truncates at some radius around $40R_{\mathrm{g}}$, transitioning into a hot flow. 
The disc inner edge is turbulent, producing intrinsic variability of the disc blackbody, whose temperature is around 0.2~keV at this truncation radius. 
This gives the low frequency `hump' peaking around 0.05~Hz in these data, and these fluctuations propagate into the hot flow, modulating its higher energy emission, so the higher energy Comptonized emission from the hot flow also has this characteristic variability feature (\citealt{Uttley_2011, Rapisarda_2016}). 
This is in addition to the turbulence generated in the hot flow itself, which has fluctuations generated at all radii 
by the MRI dynamo. 
The viscous frequency in the hotter flow should be much higher, so the turbulence generated in the flow has higher frequencies than that from the disc edge, forming the second peak in the power spectrum at $\sim 5$~Hz \citep{Rapisarda_2016}. 
There is an energy dependence in this second peak as the flow is radially stratified, emitting softer Comptonized photons close to the disc, and harder Compton closer to the black hole, where the turbulence is fastest (see Fig.~2 of K23 for a schematic of the model). 
The Comptonized emission from the hot flow illuminates the truncated disc, producing a thermal reprocessed component at low energies (in addition to the reflection/iron line emission at higher energies) which lags behind the variability of the Comptonized emission. 
This reverberation component forms a large fraction (around half) of the soft component seen in these bright hard state data as the intrinsic disc emission is quite weak due to the large truncation radius, so reprocessing from the energetically dominant Comptonized emission which peaks at $\sim 100$~keV can make a large contribution to the 
soft band ($\lesssim 2\,\si{keV}$). 
K23 showed how this model could explain the spectrum, and quantitatively fit the power spectra in different energies, and the lags between different energies as a function of frequency (see also \citealt{Kawamura_2022} for the reverberation lags).

The HIMS power spectra are quite different (Fig.~\ref{fig:ni106_psd_lag}: upper right). 
The variability at low energies is strongly suppressed (black), and the shape of the variability at high energies (blue) is much narrower than in the bright hard state, with only a single peak in the power spectrum around $\sim 5$~Hz rather than also having a second hump at lower frequencies. These two differences are easily interpreted with the same bright hard state model geometry, but where the disc now truncates very close to the black hole. 
This is clearly consistent with spectral evolution. 
The soft component in the HIMS is much hotter and brighter than in the bright hard state. 
Most of this soft emission should be intrinsic disc flux as the tail in the HIMS is weak and steep, so its contribution to reprocessing should be rather small. 
Only the inner edge of the truncated disc is turbulent in the disc, so the majority of the soft component seen in the HIMS should be stable, as observed in the power spectrum. 
We note that the abrupt drop in power at the highest frequencies seen in 0.5--1~keV (above 12~Hz) and 1--2~keV above 25~Hz is most probably due to there being a systematic overestimate of the Poisson noise level in our data. This is very small but does have an impact when the intrinsic variability is very low. 

The temperature of the inner edge of the disc is rather hard to determine as there is no clear transition seen between the disc and Comptonized emission in the HIMS spectra. 
Nonetheless, there is some sort of inflection in the spectrum at 2--3~keV which might correspond to the inner edge of the disc but this merges quite smoothly in the spectrum into the steep Comptonized tail from the hot flow. 
Thus the turbulence from the much smaller truncation radius is expected to contribute at higher energy and higher frequency than in the bright hard state, impacting the spectrum in the 2--3~keV bandpass.
However, the way the spectrum merges smoothly from the disc edge to the Comptonized emission suggests that the turbulence from these two regions might also merge, forming a single high-frequency peak in the power spectra, as observed for the highest energy emission. 

This picture where the HIMS has a much smaller hot inner flow also can explain the change in frequency of the QPO, from 0.025~Hz (bright hard state) to 5~Hz (HIMS) e.g. in the specific model where the QPOs are produced by Lense-Thirring (relativistic vertical) precession of the hot inner flow \citep{Ingram_2009}
We note that the HIMS QPO appears to be quite broad, but some of this is due to the systematic decrease of QPO frequency within the \textit{NICER} HIMS data (see \citealt{Homan_2020}). 

Thus the energy and power spectral evolution between the bright hard state and the HIMS can be interpreted as a decreasing truncation radius between the disc and hot flow. The challenge is now to explain the time lags. 
Fig.~\ref{fig:ni106_psd_lag} (lower) shows the lag between 0.5--1~keV and 1--10~keV as a function of Fourier frequency. 
The difference in lag amplitude and structure from the bright hard state to the HIMS is evident, as shown in dM21 and W21. 
The mean lag in the HIMS is around $\sim 10$~ms at high frequencies, much larger than the 1--2~ms soft lag seen in the bright low hard states, and there is no clear sign of the propagation (hard) lag which dominates at low frequencies in the bright hard state. 
The latter is easy to explain, as the long propagation lags in the bright hard state are interpreted as fluctuations generated at the (large) disc truncation radius taking time to propagate down through the (large) hot flow. 
A smaller truncation radius in the HIMS means that the inner disc edge temperature is higher, so these intrinsic fluctuations are generated at 2--3~keV not at 0.5--1~keV. 
The smaller disc truncation radius also means that the size scales for propagation are smaller, so any propagation timescale is likely shorter. 

Thus the key challenge is to explain the long negative time lags in the HIMS between 0.5--1~keV and 1--10~keV in the context of the truncated disc/hot inner flow geometry as in the bright hard state. 
With this picture, it is difficult to attribute the long soft lag to reverberation due to the small size scales.  
Instead, we explore below how this negative lag might be produced by propagation. 


\section{Modelling the variability}

\begin{figure}
	\includegraphics[trim= 0 300 0 0 , clip, width=1.\columnwidth]{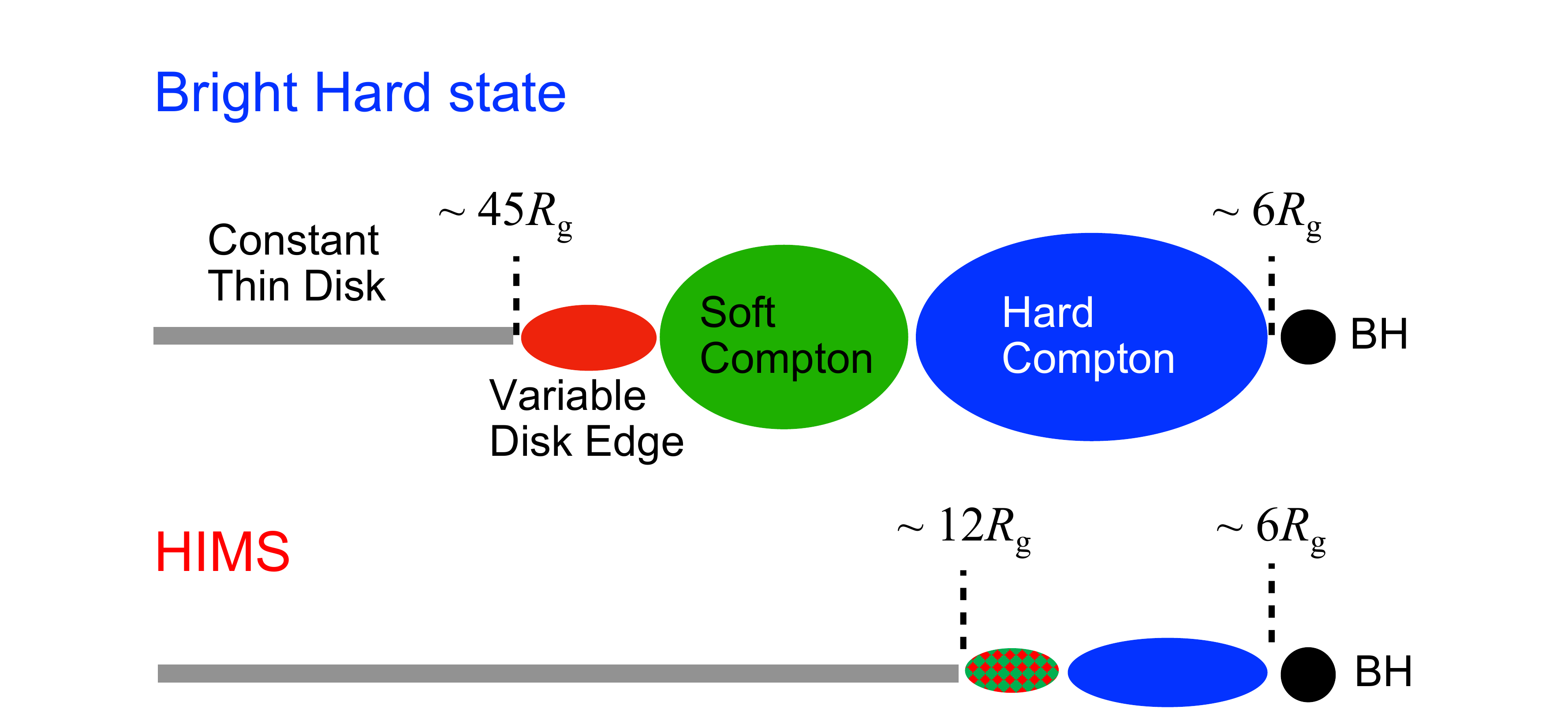}
    \caption{Schematic geometry for the bright hard state (upper, from K23) and the HIMS (lower). The disc (grey) is not variable except at radii close to where it truncates, forming a turbulent soft component on the inner disc edge (red). The Comptonised emission is radially stratified, emitting softer spectra closer to the disc where there are more seed photons (green), and harder spectra closer to the black hole (blue). The HIMS geometry is assumed to be similar, but for a much smaller transition radius, where the inner disc edge and softer Comptonisation merge (green/red hatched).}
    \label{fig:schematic}
\end{figure}

\begin{figure}
	\includegraphics[width=1.\columnwidth]{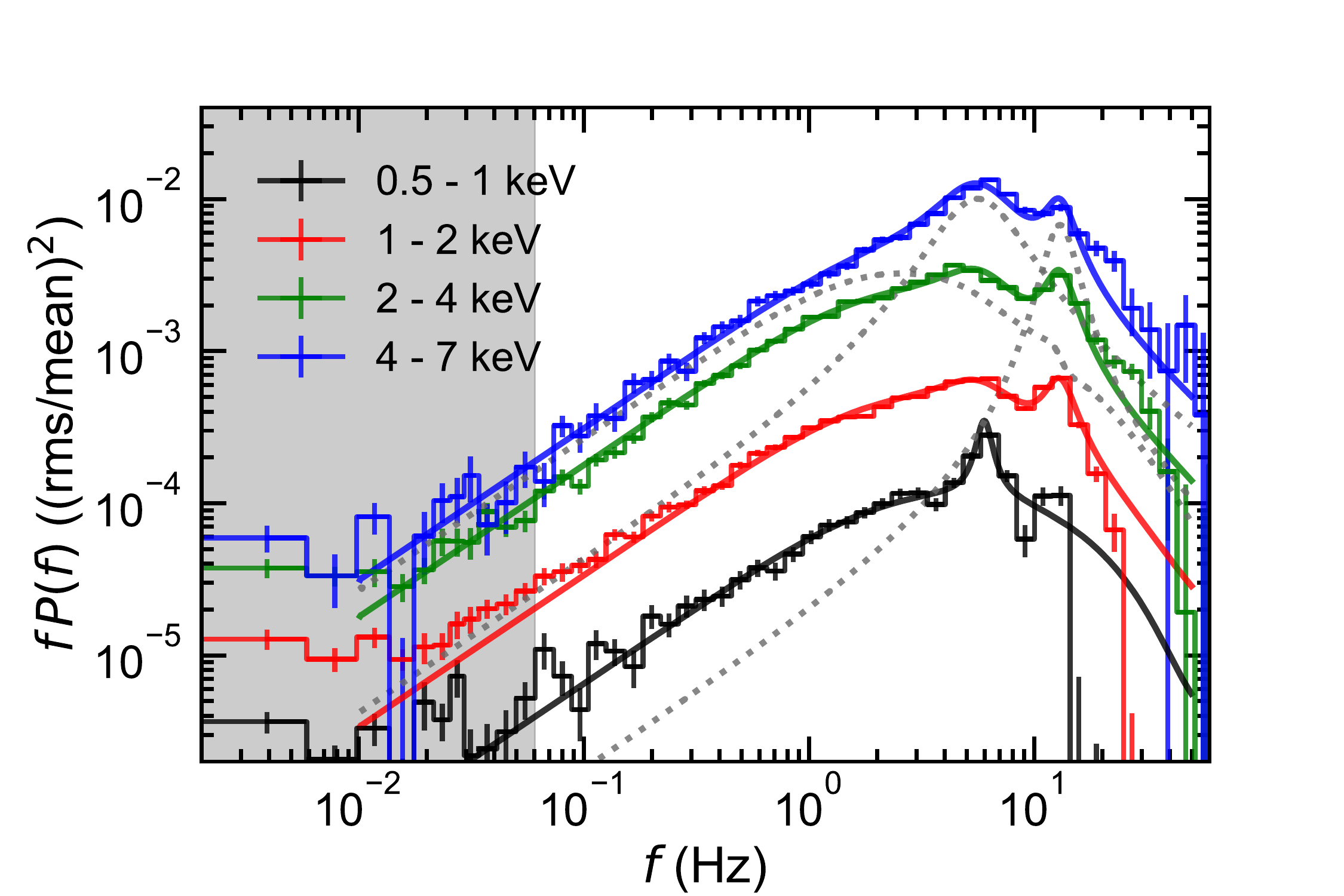}
	\includegraphics[width=1.\columnwidth]{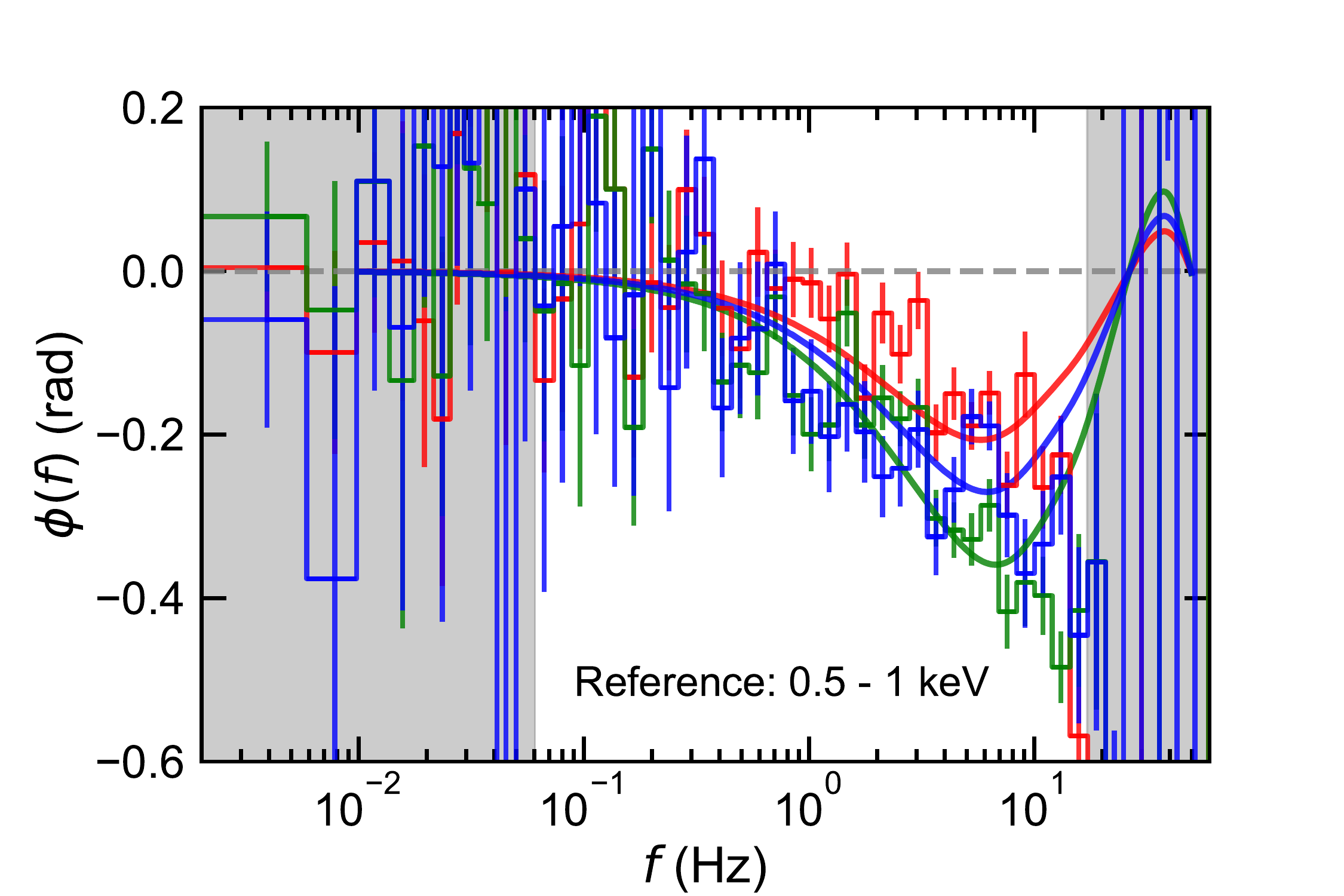}
    \caption{Simultaneous fit to the four power spectra (upper) and three phase-lag spectra (lower) with our broad-band variability model. 
    The power spectral model also includes two Lorentzian functions for the two QPOs, with these individual components shown as a grey dotted line for the highest energy band ($4\textrm{--}7\,\si{keV}$).
    The grey-shaded areas are outside the fit range, and we also ignore the power spectrum for $0.5\textrm{--}1\,\si{keV}$ above $17\,\si{Hz}$ and for $1\textrm{--}2\,\si{keV}$ above $27\,\si{Hz}$ due to likely over subtraction of the Poisson noise. 
    }
    \label{fig:psd_phase_fit}
\end{figure}

We use our previous model framework developed for the bright hard state (\citealt{Kawamura_2022}, K23). This assumes that there is a truncated disc, and a radially stratified hot flow such that it emits 
softer Comptonization at larger radii, where it sees more seed photons from the inner edge of the disc, and harder Comptonization at smaller radii, closer to the black hole (see Fig.~\ref{fig:schematic}).
K23 developed a full spectral-timing model fit formalism, but the 
HIMS spectrum is very difficult to fit with standard components such as {\sc diskbb}, especially given the very small statistical uncertainties in the \textit{NICER} data. 
We can find a fit within $\sim 1$\% of the data using three different Comptonization components, together with a broad iron line, but the parameters of these components are very difficult to constrain. 
Hence instead of doing a full spectral-timing fit, we fit only to the timing components (energy-resolved power spectra lags), and use these variability properties to constrain the contribution of each spectral component in each energy band (see Section 4 of K23). 

We assume that there are three spectral components, and that the softest is constant while the other two are variable only in normalisation. 
We associate the softest component with the stable disc blackbody, the outer variable component with the merged inner edge of the disc/soft Comptonization component, and the inner variable component with the hotter Comptonization component.
This proposed geometry for the HIMS is also shown schematically in Fig.~\ref{fig:schematic}, and the full model is detailed in the Appendix~\ref{sec:spectral_timing_model}.
We do not include reverberation in our model, so this is a clean test as to whether propagation alone can explain the long soft lags. 

The constant softest component only affects the normalization of the fractional power spectra (we use physical units of $\si{({rms/mean})^2/\si{Hz}}$). 
We do not consider spectral pivoting, i.e., the change in spectral shape, for the two variable spectral components as the much narrower bandpass of \textit{NICER} compared to the \textit{NICER} plus \textit{Insight-HXMT} data used in K23 mean that we are not so sensitive to spectral pivoting. 

We fit for the fraction of each component in each energy band, assuming that the mass accretion rate fluctuations are generated throughout the flow responsible for the two variable components, and that fluctuations generated at larger radii  propagate inwards through the rest of the flow.
Fig.~\ref{fig:psd_phase_fit} shows that this model, which uses only propagation and not reverberation, matches the broadband power spectra in different energy bands simultaneously with the phase lags (related to time lag via $2\pi f \tau(f) =\phi(f)$). We add the QPOs as separate Lorenzians (fundamental and harmonic), assuming that these are produced by only a geometric modulation of the hot flow, so they do not contribute to the phase lags. 

\begin{figure}
	\includegraphics[width=1.\columnwidth]{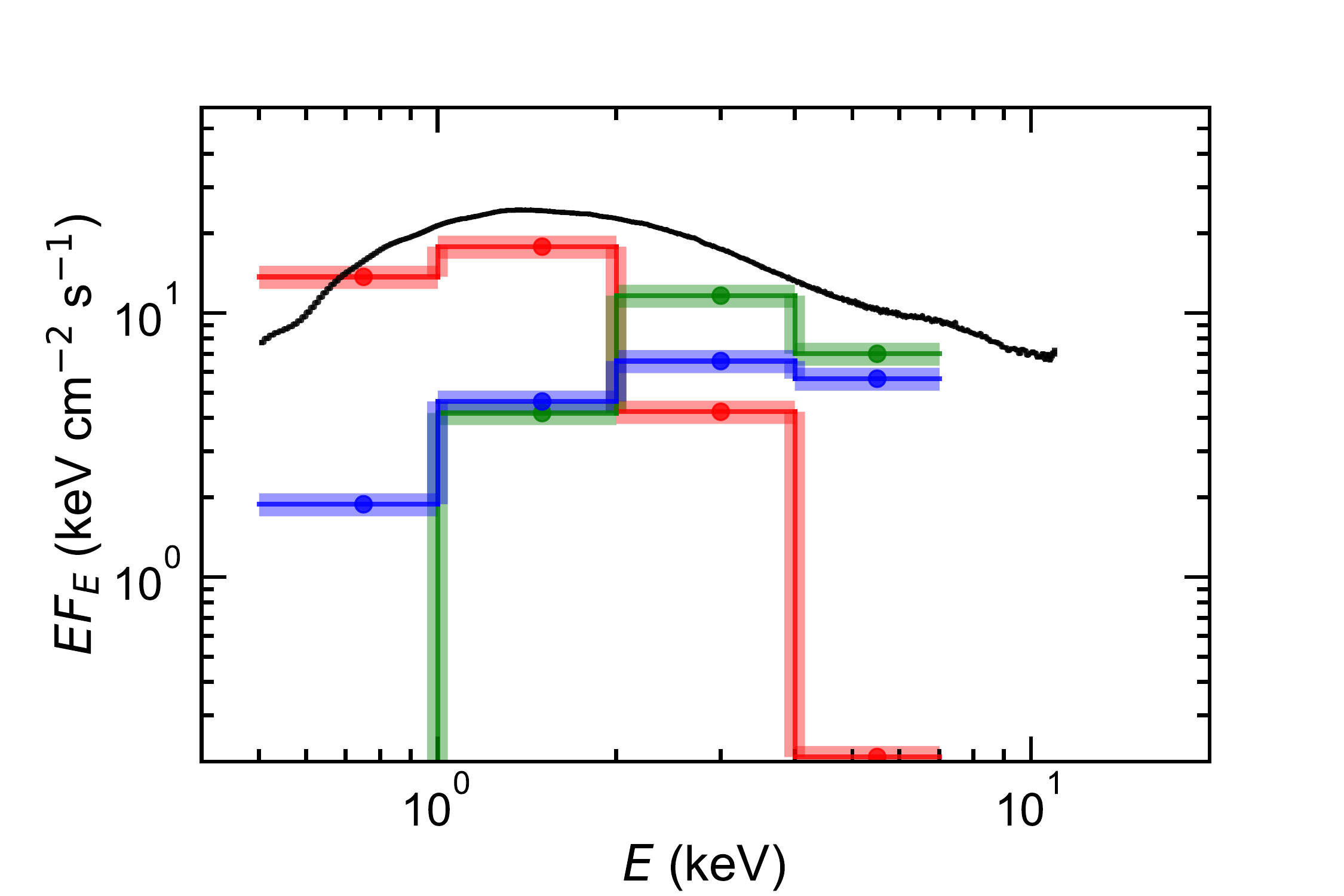}
    \caption{The black line shows the HIMS spectrum from Fig.\ref{fig:nicer_j1820_hims_spectrum}, decomposed into the shapes of the three spectral components derived from the fits to the power spectra and lags (Fig.\ref{fig:psd_phase_fit}). These are the constant soft component (red), the merged turbulent inner disc edge/soft Comptonization region  (green) and the hotter inner Comptonization region (blue). The variability is assumed to start on the turbulent inner disc edge and propagate down through the hot flow, so the blue component lags behind the green component. This gives the soft lag as this component has a broader spectrum, so it dominates the variability in the 0.5--1~keV bandpass.  
    }
    \label{fig:energy_spectra_hims_derived}
\end{figure}

We convert the fraction of each component in each energy band derived from the timing fit to produce the energy spectrum of each component (Fig.~\ref{fig:energy_spectra_hims_derived}). The stable disc (red) dominates at the lowest energies, and the soft Comptonization/turbulent inner disc edge dominates in 2--4~keV and is a rather narrow, blackbody-like component. 
The hotter Comptonization is broader: it extends to higher energies as expected but also extends down to lower energies as well so that at the softest energies it dominates over the soft Comptonization/inner disc edge component. 
This could physically arise if some fraction of the seed photons for the hotter Comptonization are from lower energies than the inner edge of the disc. These could be from the stable disc at a lower temperature than its turbulent inner edge (very likely given the dominance of this component in the spectrum), or from cyclo-synchrotron photons within the hot flow (especially if the electrons are partly non-thermal: \citealt{Malzac_2009,poutanen_2009,veledina_2013}).
Whatever their origin, this low energy extension of the hottest Compton component down below the softer emission from the turbulence at the disc edge is all that is required to make a soft lag from propagation rather than reverberation. 

\begin{figure}
	\includegraphics[width=1.\columnwidth]{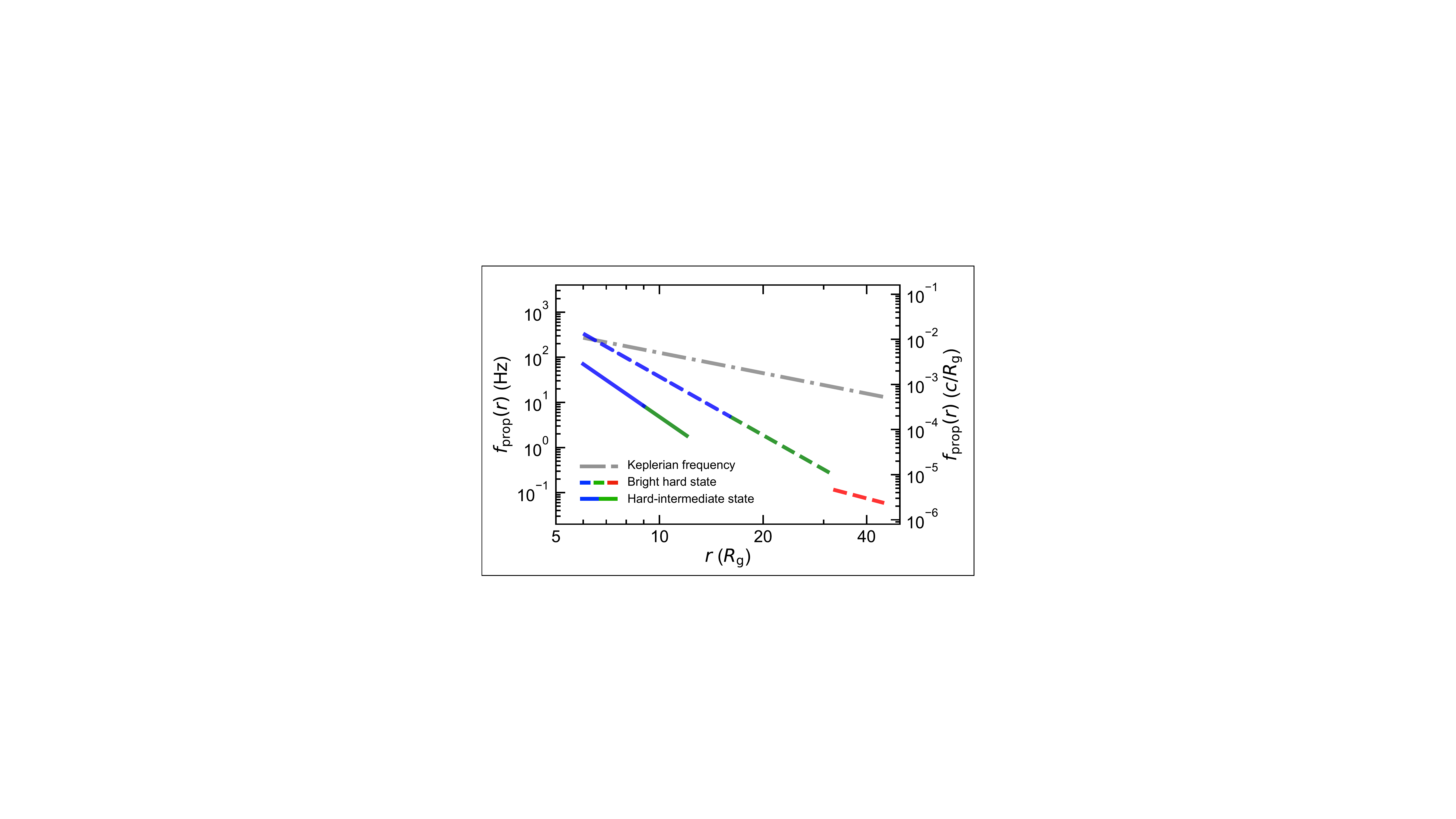}
    \caption{The propagation frequency determined from the fits to the power spectra and lags  (Fig.\ref{fig:psd_phase_fit}) for the HIMS (solid line). The dashed line shows a comparison to those derived by K23 for the bright hard state, with the grey dashed-dotted line showing the Keplarian frequency. Fitting to the HIMS requires a smaller truncation radius as expected, but also shows that the propagation speed is lower at these small radii than in the bright hard state. This might imply that the scale height of the hot flow is getting smaller, as expected from the stronger disc cooling (see also \citealt{Ingram_2012}).
    }
    \label{fig:prop_hims_lhs}
\end{figure}

Fig.~\ref{fig:prop_hims_lhs} shows the propagation frequency (left axis in Hz, right axis in $c/R_{\mathrm{g}}$) 
derived from this model of the HIMS (solid line) compared to that derived from the bright hard state (dashed line, K23), with the Keplarian orbital velocity indicated for comparison (grey dash-dotted line). 
The truncation radius in the HIMS is around $\sim 10\textrm{--}15R_g$, much smaller than the $25\textrm{--}40R_g$ truncation derived for the bright hard state. 
The propagation speed, $v_{\mathrm{p}}=rf_{\mathrm{prop}}(r)$ in units of $c$, is smaller in the HIMS hot flow than that derived for bright hard state at the same radius. This is likely indicating that the scale height of the hot flow decreases along with the decreasing inner disc radius due to the much stronger cooling of the hot flow from the much stronger intrinsic disc emission (e.g.
\citealt{Ingram_2012})


\section{Discussion and Conclusions}
\label{sec:hims}

We show that the abrupt change in lag behaviour at the HIMS/SIMS transition is quite incompatible with the proposed large change in
X-ray source scale height. Firstly there is no similarly abrupt change in hard X-ray coronal spectrum or luminosity. Secondly, the implied large change in the distance of the source from the disc should give a large change in the iron line profile from the combination of changing ionisation parameter and illumination pattern. This is not seen. Thirdly, the association of the large-scale height source with the jet ejection is incompatible with the observation that the ejection occurs after the change in lag behaviour, not before. 

Instead, in our model we propose that there is a continuous change in source properties, specifically the disc truncation radius from the bright hard state to the HIMS. The key difference in the HIMS is that the disc truncates much closer to the ISCO, so the turbulence on its inner edge merges with the (much smaller radius) turbulent hot flow, forming only a single high-frequency power spectrum peak rather than the double peak seen in the bright hard state from the much larger disc truncation radius. 

We explain the soft lag as propagation from the merged inner edge of the truncated disc/softer Comptonization region, into the hotter Comptonization region. This only requires that the spectrum of the hotter Comptonization extends down to lower energies than that of the inner disc edge/cooler Comptonization. The hotter Comptonization with its propagation lag can then dominate the variability in the 0.5--1~keV range, where the spectrum is instead dominated by the stable disc. 

Interpreting the long lag as reverberation automatically gives a large source distance, as the signal propagates at light speed. Instead, in our model, the signal is transmitted at a much slower propagation speed through the hot flow, so the observed long lag can be produced from a small source size. We show that this propagation model can quantitatively fit the power spectra and lags, and that the derived spectral component shapes are physically reasonable. Rather than requiring a huge change in X-ray source geometry, these data can be more easily interpreted as a continuation of the truncated disc/hot inner flow model to smaller truncation radii.

\section*{Acknowledgements}

We thank Megumi Shidatsu for providing the \textit{Swift} BAT data. 
This work was supported by JSPS KAKENHI Grant Numbers 18H05463 and 20H00153, and WPI MEXT. 
TK acknowledges support from JST SPRING, Grant Number JPMJFS2108. 
CD acknowledges support from the STFC consolidated grant 
ST/T000244/1, and thanks Kavli IPMU, the University of Tokyo, for support for a visit during which much of this work was completed. 

\section*{Data Availability}

All data are publically available through the HEASARC archives (\url{https://heasarc.gsfc.nasa.gov/db-perl/W3Browse/w3browse.pl}). 
The code is available from the corresponding author on reasonable request.  



\bibliographystyle{mnras}
\bibliography{bibliography.bib} 

\begin{thebibliography}{}
\makeatletter
\relax
\def\mn@urlcharsother{\let\do\@makeother \do\$\do\&\do\#\do\^\do\_\do\%\do\~}
\def\mn@doi{\begingroup\mn@urlcharsother \@ifnextchar [ {\mn@doi@}
  {\mn@doi@[]}}
\def\mn@doi@[#1]#2{\def\@tempa{#1}\ifx\@tempa\@empty \href
  {http://dx.doi.org/#2} {doi:#2}\else \href {http://dx.doi.org/#2} {#1}\fi
  \endgroup}
\def\mn@eprint#1#2{\mn@eprint@#1:#2::\@nil}
\def\mn@eprint@arXiv#1{\href {http://arxiv.org/abs/#1} {{\tt arXiv:#1}}}
\def\mn@eprint@dblp#1{\href {http://dblp.uni-trier.de/rec/bibtex/#1.xml}
  {dblp:#1}}
\def\mn@eprint@#1:#2:#3:#4\@nil{\def\@tempa {#1}\def\@tempb {#2}\def\@tempc
  {#3}\ifx \@tempc \@empty \let \@tempc \@tempb \let \@tempb \@tempa \fi \ifx
  \@tempb \@empty \def\@tempb {arXiv}\fi \@ifundefined
  {mn@eprint@\@tempb}{\@tempb:\@tempc}{\expandafter \expandafter \csname
  mn@eprint@\@tempb\endcsname \expandafter{\@tempc}}}

\bibitem[\protect\citeauthoryear{{Bright} et~al.,}{{Bright}
  et~al.}{2020}]{Bright_2020}
{Bright} J.~S.,  et~al., 2020, \mn@doi [Nature Astronomy]
  {10.1038/s41550-020-1023-5}, \href
  {https://ui.adsabs.harvard.edu/abs/2020NatAs...4..697B} {4, 697}

\bibitem[\protect\citeauthoryear{{Buisson} et~al.,}{{Buisson}
  et~al.}{2019}]{Buisson_2019}
{Buisson} D.~J.~K.,  et~al., 2019, \mn@doi [\mnras] {10.1093/mnras/stz2681},
  \href {https://ui.adsabs.harvard.edu/abs/2019MNRAS.490.1350B} {490, 1350}

\bibitem[\protect\citeauthoryear{{De Marco}, {Ponti}, {Mu{\~n}oz-Darias}  \&
  {Nandra}}{{De Marco} et~al.}{2015}]{DeMarco_2015}
{De Marco} B.,  {Ponti} G.,  {Mu{\~n}oz-Darias} T.,   {Nandra} K.,  2015,
  \mn@doi [\apj] {10.1088/0004-637X/814/1/50}, \href
  {https://ui.adsabs.harvard.edu/abs/2015ApJ...814...50D} {814, 50}

\bibitem[\protect\citeauthoryear{{De Marco}, {Zdziarski}, {Ponti}, {Migliori},
  {Belloni}, {Segovia Otero}, {Dzie{\l}ak}  \& {Lai}}{{De Marco}
  et~al.}{2021}]{DeMarco_2021}
{De Marco} B.,  {Zdziarski} A.~A.,  {Ponti} G.,  {Migliori} G.,  {Belloni}
  T.~M.,  {Segovia Otero} A.,  {Dzie{\l}ak} M.~A.,   {Lai} E.~V.,  2021,
  \mn@doi [\aap] {10.1051/0004-6361/202140567}, \href
  {https://ui.adsabs.harvard.edu/abs/2021A&A...654A..14D} {654, A14}

\bibitem[\protect\citeauthoryear{{Done}, {Gierli{\'n}ski}  \& {Kubota}}{{Done}
  et~al.}{2007}]{Done_2007}
{Done} C.,  {Gierli{\'n}ski} M.,   {Kubota} A.,  2007, \mn@doi [\aapr]
  {10.1007/s00159-007-0006-1}, \href
  {https://ui.adsabs.harvard.edu/abs/2007A&ARv..15....1D} {15, 1}

\bibitem[\protect\citeauthoryear{{Fabian}, {Rees}, {Stella}  \&
  {White}}{{Fabian} et~al.}{1989}]{Fabian_1989}
{Fabian} A.~C.,  {Rees} M.~J.,  {Stella} L.,   {White} N.~E.,  1989, \mn@doi
  [\mnras] {10.1093/mnras/238.3.729}, \href
  {https://ui.adsabs.harvard.edu/abs/1989MNRAS.238..729F} {238, 729}

\bibitem[\protect\citeauthoryear{{Fabian} et~al.,}{{Fabian}
  et~al.}{2009}]{Fabian_2009}
{Fabian} A.~C.,  et~al., 2009, \mn@doi [\nat] {10.1038/nature08007}, \href
  {https://ui.adsabs.harvard.edu/abs/2009Natur.459..540F} {459, 540}

\bibitem[\protect\citeauthoryear{{Fender}, {Homan}  \& {Belloni}}{{Fender}
  et~al.}{2009}]{fender_2009}
{Fender} R.~P.,  {Homan} J.,   {Belloni} T.~M.,  2009, \mn@doi [\mnras]
  {10.1111/j.1365-2966.2009.14841.x}, \href
  {https://ui.adsabs.harvard.edu/abs/2009MNRAS.396.1370F} {396, 1370}

\bibitem[\protect\citeauthoryear{{Homan} et~al.,}{{Homan}
  et~al.}{2020}]{Homan_2020}
{Homan} J.,  et~al., 2020, \mn@doi [\apjl] {10.3847/2041-8213/ab7932}, \href
  {https://ui.adsabs.harvard.edu/abs/2020ApJ...891L..29H} {891, L29}

\bibitem[\protect\citeauthoryear{{Ingram} \& {Done}}{{Ingram} \&
  {Done}}{2011}]{Ingram_2011}
{Ingram} A.,  {Done} C.,  2011, \mn@doi [\mnras]
  {10.1111/j.1365-2966.2011.18860.x}, \href
  {https://ui.adsabs.harvard.edu/abs/2011MNRAS.415.2323I} {415, 2323}

\bibitem[\protect\citeauthoryear{{Ingram} \& {Done}}{{Ingram} \&
  {Done}}{2012}]{Ingram_2012}
{Ingram} A.,  {Done} C.,  2012, \mn@doi [\mnras]
  {10.1111/j.1365-2966.2011.19885.x}, \href
  {https://ui.adsabs.harvard.edu/abs/2012MNRAS.419.2369I} {419, 2369}

\bibitem[\protect\citeauthoryear{{Ingram} \& {van der Klis}}{{Ingram} \& {van
  der Klis}}{2013}]{Ingram_2013}
{Ingram} A.,  {van der Klis} M.,  2013, \mn@doi [\mnras]
  {10.1093/mnras/stt1107}, \href
  {https://ui.adsabs.harvard.edu/abs/2013MNRAS.434.1476I} {434, 1476}

\bibitem[\protect\citeauthoryear{{Ingram}, {Done}  \& {Fragile}}{{Ingram}
  et~al.}{2009}]{Ingram_2009}
{Ingram} A.,  {Done} C.,   {Fragile} P.~C.,  2009, \mn@doi [\mnras]
  {10.1111/j.1745-3933.2009.00693.x}, \href
  {https://ui.adsabs.harvard.edu/abs/2009MNRAS.397L.101I} {397, L101}

\bibitem[\protect\citeauthoryear{{Kara} et~al.,}{{Kara}
  et~al.}{2019}]{Kara_2019}
{Kara} E.,  et~al., 2019, \mn@doi [\nat] {10.1038/s41586-018-0803-x}, \href
  {https://ui.adsabs.harvard.edu/abs/2019Natur.565..198K} {565, 198}

\bibitem[\protect\citeauthoryear{{Kawamura}, {Axelsson}, {Done}  \&
  {Takahashi}}{{Kawamura} et~al.}{2022}]{Kawamura_2022}
{Kawamura} T.,  {Axelsson} M.,  {Done} C.,   {Takahashi} T.,  2022, \mn@doi
  [\mnras] {10.1093/mnras/stac045}, \href
  {https://ui.adsabs.harvard.edu/abs/2022MNRAS.511..536K} {511, 536}

\bibitem[\protect\citeauthoryear{{Kawamura}, {Done}, {Axelsson}  \&
  {Takahashi}}{{Kawamura} et~al.}{2023}]{Kawamura_2023}
{Kawamura} T.,  {Done} C.,  {Axelsson} M.,   {Takahashi} T.,  2023, \mn@doi
  [\mnras] {10.1093/mnras/stad014}, \href
  {https://ui.adsabs.harvard.edu/abs/2023MNRAS.519.4434K} {519, 4434}

\bibitem[\protect\citeauthoryear{{Krawczynski} et~al.,}{{Krawczynski}
  et~al.}{2022}]{Krawczynski_2022}
{Krawczynski} H.,  et~al., 2022, \mn@doi [Science] {10.1126/science.add5399},
  \href {https://ui.adsabs.harvard.edu/abs/2022Sci...378..650K} {378, 650}

\bibitem[\protect\citeauthoryear{{Krimm} et~al.,}{{Krimm}
  et~al.}{2013}]{Krimm_2013}
{Krimm} H.~A.,  et~al., 2013, \mn@doi [\apjs] {10.1088/0067-0049/209/1/14},
  \href {https://ui.adsabs.harvard.edu/abs/2013ApJS..209...14K} {209, 14}

\bibitem[\protect\citeauthoryear{{Malzac}, {Belmont}  \& {Fabian}}{{Malzac}
  et~al.}{2009}]{Malzac_2009}
{Malzac} J.,  {Belmont} R.,   {Fabian} A.~C.,  2009, \mn@doi [\mnras]
  {10.1111/j.1365-2966.2009.15553.x}, \href
  {https://ui.adsabs.harvard.edu/abs/2009MNRAS.400.1512M} {400, 1512}

\bibitem[\protect\citeauthoryear{{Mastroserio}, {Ingram}  \& {van der
  Klis}}{{Mastroserio} et~al.}{2018}]{Mastroserio_2018}
{Mastroserio} G.,  {Ingram} A.,   {van der Klis} M.,  2018, \mn@doi [\mnras]
  {10.1093/mnras/sty075}, \href
  {https://ui.adsabs.harvard.edu/abs/2018MNRAS.475.4027M} {475, 4027}

\bibitem[\protect\citeauthoryear{{Mastroserio} et~al.,}{{Mastroserio}
  et~al.}{2021}]{Mastroserio_2021}
{Mastroserio} G.,  et~al., 2021, \mn@doi [\mnras] {10.1093/mnras/stab2056},
  \href {https://ui.adsabs.harvard.edu/abs/2021MNRAS.507...55M} {507, 55}

\bibitem[\protect\citeauthoryear{{M{\'e}ndez}, {Karpouzas}, {Garc{\'\i}a},
  {Zhang}, {Zhang}, {Belloni}  \& {Altamirano}}{{M{\'e}ndez}
  et~al.}{2022}]{Mendez_2022}
{M{\'e}ndez} M.,  {Karpouzas} K.,  {Garc{\'\i}a} F.,  {Zhang} L.,  {Zhang} Y.,
  {Belloni} T.~M.,   {Altamirano} D.,  2022, \mn@doi [Nature Astronomy]
  {10.1038/s41550-022-01617-y}, \href
  {https://ui.adsabs.harvard.edu/abs/2022NatAs...6..577M} {6, 577}

\bibitem[\protect\citeauthoryear{{Miller-Jones} et~al.,}{{Miller-Jones}
  et~al.}{2012}]{miller-jones_2012}
{Miller-Jones} J.~C.~A.,  et~al., 2012, \mn@doi [\mnras]
  {10.1111/j.1365-2966.2011.20326.x}, \href
  {https://ui.adsabs.harvard.edu/abs/2012MNRAS.421..468M} {421, 468}

\bibitem[\protect\citeauthoryear{{Miniutti} \& {Fabian}}{{Miniutti} \&
  {Fabian}}{2004}]{Miniutti_2004}
{Miniutti} G.,  {Fabian} A.~C.,  2004, \mn@doi [\mnras]
  {10.1111/j.1365-2966.2004.07611.x}, \href
  {https://ui.adsabs.harvard.edu/abs/2004MNRAS.349.1435M} {349, 1435}

\bibitem[\protect\citeauthoryear{{Miyamoto}, {Kitamoto}, {Mitsuda}  \&
  {Dotani}}{{Miyamoto} et~al.}{1988}]{Miyamoto_1988}
{Miyamoto} S.,  {Kitamoto} S.,  {Mitsuda} K.,   {Dotani} T.,  1988, \mn@doi
  [\nat] {10.1038/336450a0}, \href
  {https://ui.adsabs.harvard.edu/abs/1988Natur.336..450M} {336, 450}

\bibitem[\protect\citeauthoryear{{Nowak}, {Wilms}  \& {Dove}}{{Nowak}
  et~al.}{1999}]{Nowak_1999}
{Nowak} M.~A.,  {Wilms} J.,   {Dove} J.~B.,  1999, \mn@doi [\apj]
  {10.1086/307189}, \href
  {https://ui.adsabs.harvard.edu/abs/1999ApJ...517..355N} {517, 355}

\bibitem[\protect\citeauthoryear{{Peng} et~al.,}{{Peng}
  et~al.}{2023}]{Peng_2023}
{Peng} J.~Q.,  et~al., 2023, \mn@doi [\mnras] {10.1093/mnras/stac3238}, \href
  {https://ui.adsabs.harvard.edu/abs/2023MNRAS.518.2521P} {518, 2521}

\bibitem[\protect\citeauthoryear{{Poutanen} \& {Vurm}}{{Poutanen} \&
  {Vurm}}{2009}]{poutanen_2009}
{Poutanen} J.,  {Vurm} I.,  2009, \mn@doi [\apjl]
  {10.1088/0004-637X/690/2/L97}, \href
  {https://ui.adsabs.harvard.edu/abs/2009ApJ...690L..97P} {690, L97}

\bibitem[\protect\citeauthoryear{{Pringle}}{{Pringle}}{1981}]{Pringle_1981}
{Pringle} J.~E.,  1981, \mn@doi [\araa] {10.1146/annurev.aa.19.090181.001033},
  \href {https://ui.adsabs.harvard.edu/abs/1981ARA&A..19..137P} {19, 137}

\bibitem[\protect\citeauthoryear{{Rapisarda}, {Ingram}, {Kalamkar}  \& {van der
  Klis}}{{Rapisarda} et~al.}{2016}]{Rapisarda_2016}
{Rapisarda} S.,  {Ingram} A.,  {Kalamkar} M.,   {van der Klis} M.,  2016,
  \mn@doi [\mnras] {10.1093/mnras/stw1878}, \href
  {https://ui.adsabs.harvard.edu/abs/2016MNRAS.462.4078R} {462, 4078}

\bibitem[\protect\citeauthoryear{{Shakura} \& {Sunyaev}}{{Shakura} \&
  {Sunyaev}}{1973}]{Shakura_1973}
{Shakura} N.~I.,  {Sunyaev} R.~A.,  1973, \aap, \href
  {https://ui.adsabs.harvard.edu/abs/1973A&A....24..337S} {24, 337}

\bibitem[\protect\citeauthoryear{{Shidatsu}, {Nakahira}, {Murata}, {Adachi},
  {Kawai}, {Ueda}  \& {Negoro}}{{Shidatsu} et~al.}{2019}]{Shidatsu_2019}
{Shidatsu} M.,  {Nakahira} S.,  {Murata} K.~L.,  {Adachi} R.,  {Kawai} N.,
  {Ueda} Y.,   {Negoro} H.,  2019, \mn@doi [\apj] {10.3847/1538-4357/ab09ff},
  \href {https://ui.adsabs.harvard.edu/abs/2019ApJ...874..183S} {874, 183}

\bibitem[\protect\citeauthoryear{{Uttley}, {Wilkinson}, {Cassatella}, {Wilms},
  {Pottschmidt}, {Hanke}  \& {B{\"o}ck}}{{Uttley} et~al.}{2011}]{Uttley_2011}
{Uttley} P.,  {Wilkinson} T.,  {Cassatella} P.,  {Wilms} J.,  {Pottschmidt} K.,
   {Hanke} M.,   {B{\"o}ck} M.,  2011, \mn@doi [\mnras]
  {10.1111/j.1745-3933.2011.01056.x}, \href
  {https://ui.adsabs.harvard.edu/abs/2011MNRAS.414L..60U} {414, L60}

\bibitem[\protect\citeauthoryear{{Uttley}, {Cackett}, {Fabian}, {Kara}  \&
  {Wilkins}}{{Uttley} et~al.}{2014}]{Uttley_2014}
{Uttley} P.,  {Cackett} E.~M.,  {Fabian} A.~C.,  {Kara} E.,   {Wilkins} D.~R.,
  2014, \mn@doi [\aapr] {10.1007/s00159-014-0072-0}, \href
  {https://ui.adsabs.harvard.edu/abs/2014A&ARv..22...72U} {22, 72}

\bibitem[\protect\citeauthoryear{{Vaughan}, {Edelson}, {Warwick}  \&
  {Uttley}}{{Vaughan} et~al.}{2003}]{Vaughan_2003}
{Vaughan} S.,  {Edelson} R.,  {Warwick} R.~S.,   {Uttley} P.,  2003, \mn@doi
  [\mnras] {10.1046/j.1365-2966.2003.07042.x}, \href
  {https://ui.adsabs.harvard.edu/abs/2003MNRAS.345.1271V} {345, 1271}

\bibitem[\protect\citeauthoryear{{Veledina}, {Poutanen}  \& {Vurm}}{{Veledina}
  et~al.}{2013}]{veledina_2013}
{Veledina} A.,  {Poutanen} J.,   {Vurm} I.,  2013, \mn@doi [\mnras]
  {10.1093/mnras/stt124}, \href
  {https://ui.adsabs.harvard.edu/abs/2013MNRAS.430.3196V} {430, 3196}

\bibitem[\protect\citeauthoryear{{Wang-Ji} et~al.,}{{Wang-Ji}
  et~al.}{2018}]{Wang-ji_2018}
{Wang-Ji} J.,  et~al., 2018, \mn@doi [\apj] {10.3847/1538-4357/aaa974}, \href
  {https://ui.adsabs.harvard.edu/abs/2018ApJ...855...61W} {855, 61}

\bibitem[\protect\citeauthoryear{{Wang} et~al.,}{{Wang}
  et~al.}{2021}]{Wang_2021}
{Wang} J.,  et~al., 2021, \mn@doi [\apjl] {10.3847/2041-8213/abec79}, \href
  {https://ui.adsabs.harvard.edu/abs/2021ApJ...910L...3W} {910, L3}

\bibitem[\protect\citeauthoryear{{Wang} et~al.,}{{Wang}
  et~al.}{2022}]{Wang_2022}
{Wang} J.,  et~al., 2022, \mn@doi [\apj] {10.3847/1538-4357/ac6262}, \href
  {https://ui.adsabs.harvard.edu/abs/2022ApJ...930...18W} {930, 18}

\bibitem[\protect\citeauthoryear{{Wood} et~al.,}{{Wood}
  et~al.}{2021}]{Wood_2021}
{Wood} C.~M.,  et~al., 2021, \mn@doi [\mnras] {10.1093/mnras/stab1479}, \href
  {https://ui.adsabs.harvard.edu/abs/2021MNRAS.505.3393W} {505, 3393}

\bibitem[\protect\citeauthoryear{{Yuan} \& {Narayan}}{{Yuan} \&
  {Narayan}}{2014}]{Yuan_2014}
{Yuan} F.,  {Narayan} R.,  2014, \mn@doi [\araa]
  {10.1146/annurev-astro-082812-141003}, \href
  {https://ui.adsabs.harvard.edu/abs/2014ARA&A..52..529Y} {52, 529}

\bibitem[\protect\citeauthoryear{{Zdziarski}, {Dzie{\l}ak}, {De Marco},
  {Szanecki}  \& {Nied{\'z}wiecki}}{{Zdziarski} et~al.}{2021}]{Zdziarski_2021}
{Zdziarski} A.~A.,  {Dzie{\l}ak} M.~A.,  {De Marco} B.,  {Szanecki} M.,
  {Nied{\'z}wiecki} A.,  2021, \mn@doi [\apjl] {10.3847/2041-8213/abe7ef},
  \href {https://ui.adsabs.harvard.edu/abs/2021ApJ...909L...9Z} {909, L9}

\makeatother
\end{thebibliography}


\appendix

\section{Spectral-timing model}
\label{sec:spectral_timing_model}

The variability model used here is a simpler version of the propagating mass accretion rate fluctuation formalism of K23. 
The details are described in K22 and K23.
Here, we mainly pay attention to the differences between the variability model formalism for the HID and that of K23 for the hard state.

The disc emission is assumed to be constant here, whereas it was assumed to be variable in K23.
This comes from the properties of observed power spectra (Fig.~\ref{fig:schematic}).
Standard low/hard state power spectra are double-peaked, and so are most easily explained by a discontinuity in viscous time-scale between the inner edge of the disc and the hot flow.
However, the power spectrum in the HID here is single-peaked so we assume that the variability is dominated by the hot flow.

As a result, the disc emission does not join the propagation of mass accretion rate fluctuations and thus is not involved with the shape of power spectra and lag spectra in this paper.
It only affects the normalisation of power spectra and cross spectra in units of $(\mathrm{rms/mean})^2/\si{Hz}$ because constant emission suppresses them (\citealt{Vaughan_2003}).
The propagation of mass accretion rate fluctuations happens only within the hot flow.
In other words, the variable flow corresponds to the hot flow.
In the modelling of the propagating fluctuations, the variable flow is logarithmically split into multiple rings from $r_{\mathrm{ds}}$ to $r_{\mathrm{in}}$, instead of from $r_{\mathrm{out}}$ to $r_{\mathrm{in}}$, where $r_{\mathrm{ds}}$ is the transition radius between the disc and soft Comptonisation, $r_{\mathrm{in}}$ the inner edge of the hard Comptonisation, and $r_{\mathrm{out}}$ the outer edge of the variable disc in the hard state.
Given that the variable disc does not exist in HID, we drop the parameter $r_{\mathrm{out}}$ in this paper.

The calculation method of the model used here is almost the same as that of K23, despite the alteration of the scope of the variable flow.
The Fourier transform of the flux of energy $E$ at time $t$ is 
\begin{equation}
    X(E, f) =\sum _{n=1} ^{N_{\mathrm{r}}} w(r_n, E) \dot{M}(r_n, f),
    \label{eq:x_ft}
\end{equation}
where $\dot{M}(r_n, f)$ is the Fourier transform of the mass accretion rate of $n$th ring ($n=1, 2, \cdots, N_{\mathrm{r}}$ from outer to inner rings) centering at $r_{n}$ out of $N_{\mathrm{r}}$ rings.
We call $w(r_n, E)$ the weight and define it as 
\begin{equation}
    w(r_n, E)=\lambda (r_n)\frac{S_{0} (E, r_n)}{\dot{m}_0},
    \label{eq:weight}
\end{equation}
where $\dot{m}_{0}=1$ is the average of the normalised mass accretion rate at all the rings.
$S_{0}(E, r_n)$ is the normalised time-averaged energy spectrum at $n$th ring.
It takes the energy spectrum of the hard Comptonisation $S_{\mathrm{h}}(E)$ or that of soft Comptonisation $S_{\mathrm{s}}(E)$, depending on the location of the ring:
\begin{equation}
    S_{0} (E, r_n)=
    \begin{cases}
        S_{\mathrm{h}}(E) & (r_{\mathrm{in}} \leq r_{n} < r_{\mathrm{sh}}) ,\\
        S_{\mathrm{s}}(E) & (r_{\mathrm{sh}} \leq r_{n} < r_{\mathrm{ds}}) ,\\
    \end{cases}
\end{equation}
where $r_{\mathrm{sh}}$ is the transition radius between the soft Comptonisation and hard Comptonisation. The normalised energy spectra sum up to unity:
\begin{equation}
    \sum _{\mathrm{Y=h, s, d}} S_{Y}(E)=1,
    \label{eq:spec_sum}
\end{equation}
where $S_{\mathrm{d}}(E)$ is the energy spectrum of the constant disc.
The spectral components to be considered here are the constant disc emission, variable soft Comptonisation, and variable hard Comptonisation.
We do not include reflection components for simplicity, and we do not include spectral pivoting as the energy band over which we see the Comptonisation is small.
How much each ring contributes to the total flux is regulated by 
\begin{equation}
    \lambda=\epsilon (r_n) 2\pi r_n \Delta r_n \left/ \sum_{r_m \in \vb*{r}_{\mathrm{Y}}} \epsilon (r_m) 2\pi r_m \Delta r_m \right. ,
\end{equation}
where $\Delta r_n=r_{n} - r_{n+1}$, and $\vb*{r}_{\mathrm{Y}}$ ($\mathrm{Y}=\mathrm{h, s}$) is the collection of rings belonging to the spectral region $\mathrm{Y}$. 
$\epsilon (r_n)\propto r^{-\gamma}b(r)$ is the emissivity, where $b(r)$ is the boundary condition. 
The power spectra and cross spectra for the mass accretion rate, $(\dot{M}(r_m, t))^{*} \dot{M}(r_n, t)\,(m, n=1, 2, \cdots, N_{\mathrm{r}})$, can be calculated from the power spectrum of the intrinsic variability of the mass accretion rate 
\begin{equation}
    |A(r_n, f)|^2=\frac{2 \sigma ^2}{\pi \mu ^2} \frac{f_{\mathrm{visc}}(r_n)}{f^2 + (f_{\mathrm{visc}}(r_n))^2},
    \label{eq:mdot_power_intr}
\end{equation}
and the propagation speed $v_{\mathrm{visc}}(r)=r f_{\mathrm{visc}}(r)$ (\citealt{Ingram_2013}, K23).
Unlike K23, we do not distinguish the generator frequency and propagation frequency.
Thus, we replace both $f_{\mathrm{gen}}(r)$ and $f_{\mathrm{prop}}(r)$ in K23 with the viscous frequency $f_{\mathrm{visc}}(r)$, returning to our original model in K22 in this regard.
Since the disc emission is stable here, the viscous frequency is only for the hot flow,
\begin{equation}
    f_{\mathrm{visc}}(r)=B_{\mathrm{f}} r^{m_{\mathrm{f}}}f_{\mathrm{K}}(r)\,\,(r_{\mathrm{in}}\leq r < r_{\mathrm{ds}}),
\end{equation}
where $f_{\mathrm{K}}=(1/2\pi)r^{-3/2}$ in units of $c/R_{\mathrm{g}}$ is the Keplerian frequency
$\mu =1$ and $\sigma ^2$ are the mean and variance of the intrinsically variable mass accretion rate.
In the expression (\ref{eq:mdot_power_intr}), we employ the normalized power spectra such that their integral over positive frequency corresponds to $(\sigma/\mu)^2$.
We use the parameter $F_\mathrm{var}$ to set the variance through $\sigma / \mu = F_{\mathrm{var}}/\sqrt{N_{\mathrm{dec}}}$, where $N_{\mathrm{dec}}$ is the number of rings per radial decade and thus, $F_\mathrm{var}$ is the fractional variability per radial decade. 
Using the equations above, we can analytically calculate the power spectra $|X(E, f)|^2$ and cross spectra $(X(E_1, f))^*X(E_2, f)$ ($E_1 \neq E_2$) for the flux.

In the fitting, the spectral parameters $S_{\mathrm{Y}}(E)$ for every energy band ($E=0.5\textrm{--}1, 1\textrm{--}2, 2\textrm{--}4, 4\textrm{--}7\,\si{keV}$) are model parameters with the constraint of equation (\ref{eq:spec_sum}).
We list the parameter values used in Fig.~\ref{fig:psd_phase_fit} in Table~\ref{tab:model_parameter_values}. 

\begin{table}
\caption{Model parameter values used in Figs.~\ref{fig:ni106_psd_lag}. 
}
\begin{tabular}{ll}
\hline
Symbol&Value\\
\hline
$M_{\mathrm{BH}}$                                     &$8$      \\
$r_{\mathrm{in}}$                                     &$6$     \\
$r_{\mathrm{sh}}$                                     &$9$     \\
$r_{\mathrm{ds}}$                                     &$12$   \\
$N_{\mathrm{r}}$                                      &$40$  \\
$F_{\mathrm{var}}$                                    &$0.41$ \\
$B_{\mathrm{f}}$                                      &$226$ \\
$m_{\mathrm{f}}$                                      &$3.77$ \\
$S_\mathrm{d}(0.5\textrm{--}1\,\si{keV})$           &$0.879$   \\
$S_\mathrm{s}(0.5\textrm{--}1\,\si{keV})$           &$0$  \\
$S_\mathrm{h}(0.5\textrm{--}1\,\si{keV})$           &$1-S_{\mathrm{d}}(0.5\textrm{--}1\,\si{keV})-S_{\mathrm{s}}(0.5\textrm{--}1\,\si{keV})$   \\
$S_\mathrm{d}(1\textrm{--}2\,\si{keV})$           &$0.670$  \\
$S_\mathrm{s}(1\textrm{--}2\,\si{keV})$           &$0.157$ \\
$S_\mathrm{h}(1\textrm{--}2\,\si{keV})$           &$1-S_{\mathrm{d}}(1\textrm{--}2\,\si{keV})-S_{\mathrm{s}}(1\textrm{--}2\,\si{keV})$  \\
$S_\mathrm{d}(2\textrm{--}4\,\si{keV})$           &$0.188$     \\
$S_\mathrm{s}(2\textrm{--}4\,\si{keV})$           &$0.518$   \\
$S_\mathrm{h}(2\textrm{--}4\,\si{keV})$           &$1-S_{\mathrm{d}}(2\textrm{--}4\,\si{keV})-S_{\mathrm{s}}(2\textrm{--}4\,\si{keV})$   \\
$S_\mathrm{d}(4\textrm{--}7\,\si{keV})$           &$0.016$        \\
$S_\mathrm{s}(4\textrm{--}7\,\si{keV})$           &$0.545$    \\
$S_\mathrm{h}(4\textrm{--}7\,\si{keV})$           &$1-S_{\mathrm{d}}(4\textrm{--}7\,\si{keV})-S_{\mathrm{s}}(4\textrm{--}7\,\si{keV})$  \\
\hline
\end{tabular}\\
\label{tab:model_parameter_values}
\end{table}

\bsp	
\label{lastpage}
\end{document}